\documentclass{article}

     \usepackage[final]{neurips_2020}

\usepackage[utf8]{inputenc} 
\usepackage[T1]{fontenc}    
\usepackage{hyperref}       
\usepackage{url}            
\usepackage{booktabs}       
\usepackage{amsfonts}       
\usepackage{nicefrac}       
\usepackage{microtype}      
\usepackage{graphicx}
\usepackage[english]{babel}
\usepackage[utf8]{inputenc}

\usepackage[nottoc]{tocbibind}

\title{Self-Supervised Transformers for Activity Classification using Ambient Sensors}

\author{
    Luke Hicks\thanks{hicksl@coventry.ac.uk}, Vasile Palade\thanks{vasile.palade@coventry.ac.uk},
    Ibrahim Almakky\thanks{ibrahimalmakky@gmail.com} \\
    Coventry University\\
    Coventry, United Kingdom\\
    \And
    Ariel Ruiz-Garcia\thanks{ariel.9arcia@gmail.com} \\
    LatinX in AI and SeeChange.ai\\
    Manchester, United Kingdom\\
}

\begin{document}
 
\maketitle
  
\section{Introduction}
Providing care for ageing populations is an onerous task, and as life expectancy estimates continue to rise, the number of people that require senior care is growing rapidly \cite{OfficeforNationalStatistics2018}. Ambient sensing enables data collection within healthcare facilities where it essential to be non-intrusive, but also escalates the complexity in performing human activity classification. This paper proposes a methodology based on Transformer Neural Networks to classify the activities of a resident within an ambient sensor based environment. We also propose a methodology to pre-train Transformers in a self-supervised manner, as a hybrid autoencoder-classifier model instead of using contrastive loss. The social impact of the research is considered with wider benefits of the approach and next steps for identifying transitions in human behaviour. In recent years there has been an increasing drive for integrating sensor based technologies within care facilities for data collection. This allows for employing machine learning for many aspects including activity recognition and anomaly detection. Due to the sensitivity of healthcare environments, some methods of data collection used in current research are considered to be intrusive within the senior care industry, including cameras for image based activity recognition, and wearables for activity tracking, but recent studies have shown that using these methods commonly result in poor data quality due to the lack of resident interest in participating in data gathering. This has led to a focus on ambient sensors, such as binary PIR motion, connected domestic appliances, and electricity and water metering. By having consistency in ambient data collection, the quality of data is considerably more reliable, presenting the opportunity to perform classification with enhanced accuracy. Therefore, in this research we looked to find an optimal way of using deep learning to classify human activity with ambient sensor data.

\section{Methodology}

Transformer neural networks have become the predominant choice for Natural Language Processing (NLP) related tasks, largely due to their non-sequential processing ability, long term memory and multi-attention mechanisms. However, they have yet to prove their ability to perform well in other domains such as classification or even time-series prediction. We propose using Transformers for activity classification in Ambient Sensor environments. For this task, some modifications to the Transformer model are required. 




Transformers, like other recurrent models, are composed of an encoder and decoder model. Because we are not interested in predicting sequences, we propose using the Encoder element of the transformer only. Furthermore, because we do not embed time onto the model input, we remove the embedding layer and replace it with a fully connected layer. A Softmax MLP is also attached as a head instead of the decoder. 

The dataset used in this work is a collection from a study into activities of daily living in older adulthood and dementia \cite{Cook}. An apartment is fitted with multiple sensors, and accommodated by an individual resident who is monitored by a researcher for activity labelling. The sensors used include motion, kitchen appliances and consumable items, door states, hot and cold water usage, temperature readings and apartment electricity usage. The data is marked at intervals based on the start and the end of which activity is being performed. This data is presented to the model as a one-hot vector of motion sensor activations. 

\subsection{Self-Supervised Learning}

\begin{figure*}[!htbtp]
        \centering
		\includegraphics[width=10cm]{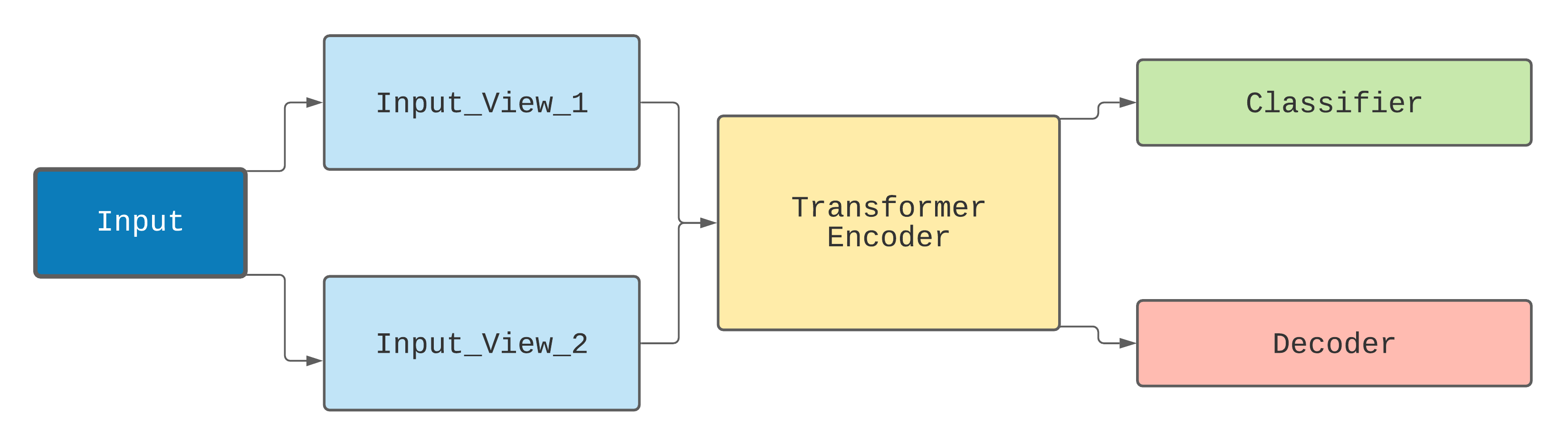}
		\caption[self-supervised model]{Visualisation of the self-supervised training approach proposed in this work. } 
		\label{fig:model}%
\end{figure*}	
	
Taking into account the proven ability of self-supervised learning in boosting the performance of models where limited labelled data is available, we propose using self-supervised learning on unlabelled data for the transformer model. Self-supervised learning uses a contrastive loss to determine whether two samples are different views from the same sample or from different samples. Instead of using contrastive loss as empirical self-supervised models, we propose designing the model as a hybrid autoencoder and binary classifier model. 

Recall that the recognition transformer model described earlier is made up of an encoder and classifier, we propose making the classifier a binary classifier and attaching an auxiliary decoder head to the model to create an autoencoder. Autoencoders are composed of a function $f$ that encodes an input $x$ into a hidden representation $h$, which is mapped back to a representation $y$ of the input $x$ through a decoder function $g$. They are formally defined as: $y = g(f(x))$. However, since in self-supervised learning a batch is traditionally made up of two different views $x_i$ and $x_j$ of the same input $x$, thus resulting in $2N$ data points, instead of using $x_i$ and $x_j$ as the target reconstruction, we use $x$. In effect, this tells the model to reconstruct both views of the input as the original input which means the model will never converge. However, we just need the model to learn to associate both inputs rather than produce perfect reconstructions. 

Let $L_{a}$ be an $l_2$ distance between the input $x$ and the reconstruction $y$, let $L_{c}$ be a binary classification loss, then the loss of the proposed model is formally defined as $L = \gamma * L_{a} + L_{c}$ where $\gamma$ is a scale factor and $0 \leq \gamma \leq $1. The binary classifier in this formulation tells us whether the two samples are two different views of the same input or not. Therefore, the output of the encoder is randomised in such a way that half of the pairs fed to the classifier are from the same input and the other half randomised. These are then concatenated and fed to the classification head as a single feature vector.

By training using a classification, loss we allow the model to learn more sparse representations useful for classification, which is our primary goal. The regression loss helps in retaining contextual information, which in our case is important for recognition. Once training is done, the decoder is discarded and the rest of the model is used to initialise the transformer encoder.

To the best of the authors' knowledge this is the first approach that formulates self-supervised learning as a hybrid autoencoder and binary classifier approach. This approach can also be applied to other models like CNN.  

\section{Broader Social Impact}
Research will continue in this area to seek how this method of classification can assist in the early detection of health decline. By recognising transitions in behaviour, carers and medical professionals can detect signs of illness, enabling automation processes for informing carers and improving the level of care being provided. Furthermore, the methodology described in this work will be extended for prediction, i.e. predict when a specific event will occur, in addition to recognition. This would in effect allow us to predict and prevent accidents and incidents in home cares. 

\bibliographystyle{unsrt}
\nocite{*}
\bibliography{references}
\end{document}